\documentclass[twocolumn]{aastex701}

\usepackage{newtxtext}          
\usepackage{newtxmath}
\usepackage{amsmath}	        
\usepackage{textcomp,gensymb}	
\usepackage{xcolor}             

\newcommand{\thisstar}{HD~715}
\newcommand{\thisstarb}{HD~715~b}

\newcommand\ms{\ensuremath{\text{m}\,\text{s}^{-1}}}
\newcommand\cmssq{\ensuremath{\text{cm}\,\text{s}^{-2}}}
\newcommand\masyr{\ensuremath{\text{mas}\,\text{yr}^{-1}}}

\received{2026 April 13}
\revised{2026 July 20}
\accepted{2026 July 23}

\begin{document}

\title{HD 715 b, a Warm Jupiter Orbiting a Massive Subgiant Discovered from a Single TESS Transit\footnote{This paper includes data gathered with the 6.5 meter Magellan Telescopes located at Las Campanas Observatory, Chile.}}

\correspondingauthor{Alexander Venner}

\author[0000-0002-8400-1646]{Alexander Venner}
\affiliation{Max Planck Institute for Astronomy, 69117 Heidelberg, Germany}
\affiliation{Centre for Astrophysics, University of Southern Queensland, Toowoomba, QLD 4350, Australia}
\email[show]{alvenner@mpia.de}

\author[0000-0001-9957-9304]{Robert A. Wittenmyer}
\affiliation{Centre for Astrophysics, University of Southern Queensland, Toowoomba, QLD 4350, Australia}
\email{Rob.Wittenmyer@unisq.edu.au}

\author[0000-0001-7246-5438]{Andrew Vanderburg}
\affiliation{Center for Astrophysics, Harvard \& Smithsonian, Cambridge, MA 02138, USA}
\email{avanderburg@cfa.harvard.edu}

\author[0000-0001-6588-9574]{Karen A. Collins}
\affiliation{Center for Astrophysics, Harvard \& Smithsonian, Cambridge, MA 02138, USA}
\email{karen.collins@cfa.harvard.edu}

\author[0000-0001-9911-7388]{David W. Latham}
\affiliation{Center for Astrophysics, Harvard \& Smithsonian, Cambridge, MA 02138, USA}
\email{dlatham@cfa.harvard.edu}

\author[0000-0001-6637-5401]{Allyson Bieryla}
\affiliation{Center for Astrophysics, Harvard \& Smithsonian, Cambridge, MA 02138, USA}
\affiliation{Centre for Astrophysics, University of Southern Queensland, Toowoomba, QLD 4350, Australia}
\email{abieryla@cfa.harvard.edu}

\author[0000-0002-0692-7822]{Tyler R. Fairnington}
\affiliation{Department of Astronomy \& Astrophysics, University of Chicago, Chicago, IL 60637, USA}
\affiliation{Centre for Astrophysics, University of Southern Queensland, Toowoomba, QLD 4350, Australia}
\email{tfairnington@uchicago.edu}

\author[0000-0003-0918-7484]{Chelsea X. Huang}
\affiliation{Centre for Astrophysics, University of Southern Queensland, Toowoomba, QLD 4350, Australia}
\email{Chelsea.Huang@unisq.edu.au}

\author[0000-0001-5222-4661]{Timothy R. Bedding}
\affiliation{Sydney Institute for Astronomy (SIfA), School of Physics, University of Sydney, NSW 2006, Australia}
\email{tim.bedding@sydney.edu.au}

\author[0000-0003-4976-9980]{Samuel Grunblatt}
\affiliation{Department of Physics and Astronomy, University of Alabama, Tuscaloosa, AL 35487-0324, USA}
\email{sgrunblatt@ua.edu}

\author[0000-0001-9827-1463]{Sydney Jenkins}
\affiliation{Department of Physics and Kavli Institute for Astrophysics and Space Research, Massachusetts Institute of Technology, Cambridge, MA 02139, USA}
\email{sydneyaj@mit.edu}


\author[0000-0003-1305-3761]{R. Paul Butler}
\affiliation{Earth and Planets Laboratory, Carnegie Institution for Science, Washington, DC 20015}
\email{pbutler@carnegiescience.edu}

\author[0000-0002-5226-787X]{Jeffrey D. Crane}
\affiliation{The Observatories of the Carnegie Institution for Science, Pasadena, CA 91101, USA}
\email{crane@carnegiescience.edu}

\author[0000-0001-9261-8366]{Jhon Yana Galarza}
\affiliation{Departamento de Astronomía, Universidad de Concepción, Casilla 160-C, Concepción, Chile}
\email{jyanag@udec.cl}

\author[0000-0002-1160-7970]{Jonathan Horner}
\affiliation{Centre for Astrophysics, University of Southern Queensland, Toowoomba, QLD 4350, Australia}
\email{Jonti.Horner@unisq.edu.au}

\author[0000-0002-2532-2853]{Steve B. Howell}
\affiliation{NASA Ames Research Center, Moffett Field, CA 94035, USA}
\email{steve.b.howell@nasa.gov}

\author[]{Marc Huten}
\affiliation{Citizen Scientist, Zooniverse c/o University of Oxford, Keble Road, Oxford OX1 3RH, UK}
\email{huten.marc@orange.fr}

\author[0000-0002-7084-0529]{Stephen R. Kane}
\affiliation{Department of Earth and Planetary Sciences, University of California, Riverside, CA 92521, USA}
\email{skane@ucr.edu}

\author[0000-0003-0412-9664]{David Osip}
\affiliation{Las Campanas Observatory, Carnegie Institution for Science, Casilla 601 La Serena, Chile}
\email{dosip@carnegiescience.edu}

\author[0009-0004-8891-4057]{Marcelo Said Tala Pinto}
\affiliation{Department of Astronomy, Ohio State Univesity, McPherson Laboratory, Columbus, OH 43210, USA}
\email{tala.1@osu.edu}

\author[0000-0002-8964-8377]{Samuel N. Quinn}
\affiliation{Center for Astrophysics, Harvard \& Smithsonian, Cambridge, MA 02138, USA}
\email{squinn@cfa.harvard.edu}

\author[0009-0008-2801-5040]{Johanna Teske}
\affiliation{Earth and Planets Laboratory, Carnegie Institution for Science, Washington, DC 20015}
\affiliation{The Observatories of the Carnegie Institution for Science, Pasadena, CA 91101, USA}
\email{jteske@carnegiescience.edu}

\author[0000-0002-8681-6136]{Steve Shectman}
\affiliation{The Observatories of the Carnegie Institution for Science, Pasadena, CA 91101, USA}
\email{shec@carnegiescience.edu}

\author[0000-0003-2527-1475]{Shreyas Vissapragada}
\affiliation{The Observatories of the Carnegie Institution for Science, Pasadena, CA 91101, USA}
\email{svissapragada@carnegiescience.edu}

\author[0000-0001-7294-5386]{Duncan Wright}
\affiliation{Centre for Astrophysics, University of Southern Queensland, Toowoomba, QLD 4350, Australia}
\email{duncan.wright@unisq.edu.au}

\author[0000-0002-4891-3517]{George Zhou}
\affiliation{Centre for Astrophysics, University of Southern Queensland, Toowoomba, QLD 4350, Australia}
\email{George.Zhou@unisq.edu.au}

\begin{abstract}

Transiting Warm Jupiters orbiting bright stars are scarcely known, but a substantial number of these planets have been uncovered by the TESS mission. In this work we report the discovery and confirmation of a transiting Warm Jupiter orbiting the bright ($V$ = 8.70) subgiant star HD~715 (TOI-6893). HD~715~b was originally identified from a single transit event in TESS Sector~70. We acquired two seasons of radial velocity follow-up using the PFS, TRES, and MINERVA-Australis spectrographs, which clearly detect the planetary orbit. At $M_*=1.60\pm0.04~M_\odot$, HD~715 thus becomes one of the most massive stars known to host a transiting Warm Jupiter. From a joint fit to the TESS transit and our RVs, we find that HD~715~b has a $56.40^{+0.19}_{-0.23}$~day orbital period and measure a precise mass and radius of $0.635\pm0.056~M_J$ and $1.07\pm0.04~R_J$, rendering it a relatively typical member of the Warm Jupiter population. The moderate orbital eccentricity of $0.21^{+0.07}_{-0.08}$ contrasts with the higher eccentricities of other similar Warm Jupiters orbiting evolved stars, indicating a greater diversity in orbital eccentricities than previously known. The absence of additional transits in preceding TESS observations in Sectors~42 and 43 allows us to further refine the range of possible orbital periods to $[56.01, 56.35]$~days. Though the ephemeris uncertainty remains substantial, future photometric observations may nonetheless allow for recovery of the transit of HD~715~b.


\end{abstract}




\section{Introduction} \label{sec:intro}

Giant planets with significantly shorter orbital periods than seen in the solar system comprise a major section of the known population of exoplanets. Our knowledge of these short-period giant planets is dominated by the iconic Hot Jupiters with orbital periods below $<$10~days. However, the ``Warm Jupiters" found at longer orbital periods are less well-known. These planets are conventionally thought to be less frequent than Hot Jupiters, with a low-occurrence ``period valley" prevailing between $\approx$10 -- 100 days \citep{Udry2003, Santerne2016}. Nonetheless, transit surveys have detected an increasingly large number of Warm Jupiters, leading to their growing recognition as an independent class of exoplanets.\footnote{For the purposes of this work, we define a ``Warm Jupiter" as a planet with mass $M_p>0.1~M_J$, radius $R_p>8~R_\oplus$ and orbital period $10<P<100$~days. Varying these limits does not meaningfully alter our conclusions.}

Several lines of evidence suggest that Warm Jupiters and Hot Jupiters have different formation histories. The distribution of Warm Jupiter orbital eccentricities is inconsistent with dominance of high-eccentricity migration \citep{Petrovich2016}; their obliquities are preferentially more aligned than Hot Jupiters \citep{Rice2022}; and they are frequently found flanked by smaller planets \citep{Huang2016} -- dissimilar to the situation for Hot Jupiters \citep{Sha2026}.

Much of the current research into transiting exoplanets has been shaped by data collected by Transiting Exoplanet Survey Satellite mission \citep[TESS;][]{TESS}, and the same is true for Warm Jupiters. Transits of giant planets across Sun-like stars typically produce prominent transit events ($\sim$1\% in depth), and as it performs its all-sky survey of bright stars, TESS has detected scores of Warm Jupiters \citep{Dong2021.catalogue}. TESS observes in discrete $\sim$26-day sectors, meaning that most stars are only observed for a few weeks at a time. While this is disadvantageous for detection of longer-period giant planets, many of the most interesting such exoplanets known have in fact been discovered by the mission, often on the basis of small numbers of TESS transits \citep[e.g.][]{Dong2021, Dalba2022, Zakhozhay2022, Hobson2023, Mann2023, Gupta2024, Essack2025, Bieryla2026, Rodel2026}. Thus while TESS has now observed almost the entire sky in at least one sector, as observations continue and stars are continually re-observed, new Warm Jupiters are likely to be revealed.

A related area of exoplanet research where TESS has made significant contributions is the study of giant exoplanets orbiting evolved stars. As stars more massive than the Sun depart from the main sequence they will cool and expand, which has the effect of making orbiting planets more likely to transit due to the larger stellar radius, and easier to study since hot main sequence stars are otherwise challenging for radial velocity (RV) follow-up due to their typically broadened spectral lines. Transiting giant planets orbiting evolved ``retired" A/F-type stars therefore provide unique information about the planetary systems of massive stars \citep[e.g.][]{Grunblatt2016, Grunblatt2019, Rodriguez2021}. Warm Jupiters are particularly amenable to detection by TESS, even from single transits, due to their large transit depths and durations. TESS discoveries have therefore provided a great deal of insight into the late stages of Warm Jupiter evolution \citep[e.g.][]{Nielsen2019, Brahm2019, Huber2019, Wittenmyer2022, Grunblatt2023, Chontos2024}.

In this work we present the discovery and confirmation of a transiting Warm Jupiter, \thisstarb{}, orbiting a bright ($V=8.7$) and comparatively massive ($M_*=1.60\pm0.04~M_\odot$) subgiant star. The planet was initially detected from a single transit observed by TESS, and confirmed and characterised through RV follow-up using the MINERVA-Australis, PFS, and TRES spectrographs. In Section~\ref{sec:observations} we detail the observations involved in this discovery, and our characterisation of the system properties in Section~\ref{sec:analysis}. In Section~\ref{sec:results} we present the results of our joint model to the transit and RVs, and conclude with a discussion in Section~\ref{sec:discussion}.

\section{Observations} \label{sec:observations}

\subsection{TESS photometry} \label{subsec:TESS}

\begin{figure*}
    \centering
    \includegraphics[width=\textwidth]{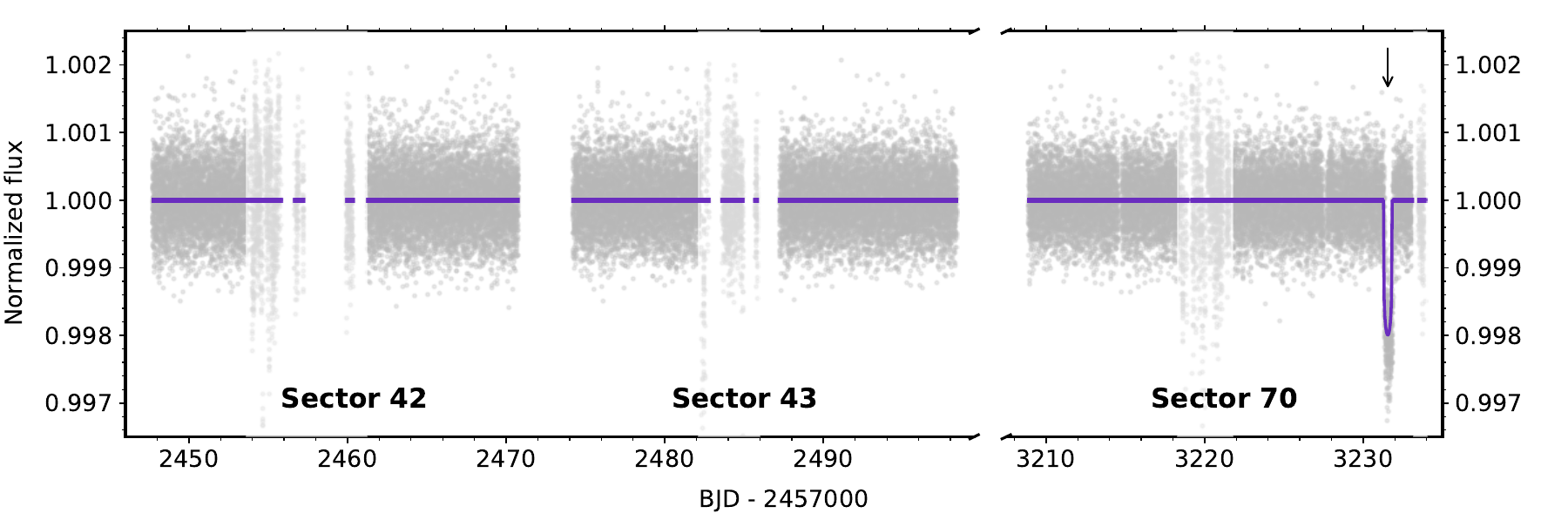}
    \caption{TESS photometry of \thisstar{} (TIC~272125414, TOI-6893). We show the normalised and flattened 2-minute cadence photometry as grey points, including mid-sector observations affected by scattered light in light colours (see text). The location of the single transit detected in Sector~70, at TESS~BJD 3231.55, is highlighted with an arrow.
    }
    \label{fig:lightcurve}
\end{figure*}

Our study of \thisstar{} (TIC~272125414) was instigated by the detection of a transit during TESS observations. It is notable in this context that TESS observations of the star began comparatively late in the mission; this is because it lies near the ecliptic plane (ecliptic latitude = +3.22$\degree$), and due to the ecliptic-averse observing strategy employed by TESS \citep{TESS} the star was not observed during the first three years of the mission.

\thisstar{} was instead first observed by TESS during the first round of ecliptic plane observations in Sectors~42 and 43 (2021-08-20 -- 2021-10-12). Observations in these sectors were strongly affected by crossings of the Earth and Moon into the TESS field-of-view (\citealt{TESS_DRN_S42, TESS_DRN_S43}; see also \citealt{DholakiaPalethorpe2024}). As a result the corresponding TESS datasets for \thisstar{} contain unusually large amounts of scattered light in mid-sector observations. The processed data reveal a largely constant flux with no evidence for transits, including during mid-sector observations.

\thisstar{} was then not observed again until TESS Sector~70 (2023-09-20 -- 2023-10-16). During this sector the Moon passed directly through the field of view near to the middle of the sector \citep{TESS_DRN_S70}, resulting in another scattered light event in the mid-sector photometry. Near to the end of the Sector~70 observations, a clear transit is visible at TESS~BJD 3231.55\footnote{i.e. BJD - 2457000.} with a depth of $\sim$1600~ppm and a duration of $\sim$13.4~hours. The transit was first identified through the Planet Hunters TESS citizen science project \citep{Eisner2020},\footnote{\url{https://www.zooniverse.org/projects/nora-dot-eisner/planet-hunters-tess/talk/2112/3184522}} and posted as an ExoFOP Community TESS Object of Interest (CTOI), on 2023-12-01. The depth of the transit nominally corresponds to a companion radius of $\sim$1.0~$R_J$, compatible with a giant planet. The transit was subsequently upgraded to a TESS Object of Interest (TOI) on 2024-02-01, with the designation of TOI-6893. We produce a systematics-corrected light curve from the three sectors of TESSobservations following \citet{Vanderburg2019}, which is shown in Figure~\ref{fig:lightcurve}.

Given that TESS observed only one transit in three sectors of observations, the orbital period of the planet candidate cannot be directly constrained from the photometry. However, the transit duration $T_D$ can be derived in first order from the stellar density, the impact parameter of the transit, and the companion orbital period, eccentricity, and argument of periastron. Given that the stellar density can be externally constrained, priors can be applied to the orbital eccentricity, and the impact parameter can be constrained by the transit shape, then by constructing an appropriate transit model it becomes possible to estimate the orbital period of a companion from a single transit \citep[hence][]{Yee2008, Osborn2016, Sandford2019}.

To obtain a preliminary estimate for the orbital period of the \thisstar{} planet candidate, we use the single-transit model implemented in \citet{Venner2026}. This runs a conventional transit model generated by the \texttt{batman} package \citep{batman}, applying the single transit period probability prior from \citet{Kipping2018}. Using our constraint on the stellar density (see Section~\ref{subsec:star}) and the provisional assumption of zero orbital eccentricity, we estimate $P=55^{+28}_{-17}$~d. This may appear surprisingly short given the three sectors of TESS observations for the star, however due to the aforementioned gaps in the actual TESS coverage there are many ``blind spots" beyond $P\gtrsim30$~d where a second transit could have been missed (on which see Section~\ref{subsec:period_refinement}). We therefore consider this to be a plausible guess for the orbital period of the planet candidate.

As \thisstar{} is not scheduled to be observed by TESS again for the foreseeable future, to resolve the orbital period as required for confirmation of the planet candidate would wholly depend on follow-up observations. Given $P=55^{+28}_{-17}$~d, it appeared likely that the orbital period would be short compared to an observing season, and assuming a Jupiter-like mass the planet candidate would generate a reflex RV signal at the level of tens of \ms{}, which appeared likely to be detectable given the brightness of the star. This motivated us to begin RV observations of the target.

\subsection{Spectroscopic Observations}

\subsubsection{MINERVA-Australis}

Due to the location of \thisstar{} on the ecliptic, the star passes behind the Sun and is correspondingly unobservable between approximately February and May. This delayed the onset of ground-based follow-up of the star following its announcement as a TOI.

We acquired our first spectroscopic observations of \thisstar{} with the MINERVA-Australis spectrograph array at Mount Kent Observatory, Queensland, Australia \citep{Addison2019}. Our first MINERVA-Australis spectrum for \thisstar{} was collected on 2024-05-22, which showed the spectrum of a subgiant star with a moderately high $v\sin i (\approx$11~k\ms{}). Between 2024-05-22 and 2024-08-31 we accumulated a total of 17 MINERVA-Australis RVs, covering approximately two orbital revolutions of \thisstarb{} and attaining a median precision of 31~\ms{}. By this time it had become clear from our PFS observations that the RV semi-amplitude $K$ of the transiting planet \thisstarb{} was smaller than the MINERVA-Australis RV precision. We therefore chose to halt MINERVA-Australis observations of the star. We present this dataset in Appendix~\ref{appendix:RV}.

Unlike the other RV data used in this work, the MINERVA-Australis observations provide absolute values for the stellar RV. The barycentric system RV derived from these observations is $+11.01\pm0.01$~k\ms{}. This nominally differs from the \textit{Gaia}~DR3 value for the stellar RV, $+11.99\pm0.18$~k\ms{} \citep{GaiaDR3}. However, there is a $-0.783$~k\ms{} offset between the MINERVA-Australis and \textit{Gaia} RV zero-points \citep{Wittenmyer2026}, and after accounting for this the renormalised RV of $+11.79\pm0.02$~k\ms{} shows stronger agreement with the \textit{Gaia}~DR3 RV. We adopt the latter value for the barycentric RV for the system.

\subsubsection{PFS} \label{subsec:PFS}

Throughout the 2024 and 2025 observing seasons we followed up \thisstar{} with the Planet Finder Spectrograph \citep[PFS;][]{Crane2006, Crane2008, Crane2010} on the 6.5~m Magellan II telescope at Las Campanas Observatory, Chile. Observations in semesters 2024A and 2024B were acquired using left over time from a PFS observing program targeting another similar Warm Jupiter system (A. Venner et al., in prep.). 
In 2025, we continued observing \thisstar{} through a third observing programme focusing on TESS Warm Jupiters (PI~Collins).

PFS is mounted on-telescope for $\sim$1-week observing runs that are spaced temporally by irregular intervals, typically every $\approx$1 -- 2 months. In each run we attempted to make observations on two nights, once near the start and once near the end of each run, in order to maximise the capture of orbital motion. Observations were executed as 900~s exposures, typically two exposures observed sequentially in order to average out the effects of short-term stellar variability, though on some nights only one exposure was obtained.

We obtained our first PFS epoch for \thisstar{} on 2024-05-29, and observed on a total of 10 nights in 2024 up to 2024-12-12. In 2025, we acquired observations across a further 6 nights between 2025-06-10 and 2025-10-28. To extract the RVs, we used the standard PFS reduction pipeline to reduce the spectra \citep{Butler1996}. After nightly binning, we are left with 16 RVs with a median precision of 2.1~\ms{} that show clear periodic variability. This data is tabulated in Appendix~\ref{appendix:RV}.

\subsubsection{TRES}

By mid-2024 we had detected significant RV variability in our PFS observations that could be attributed to the transiting planet candidate. However, the intermittent cadence of PFS observations left the orbital eccentricity poorly constrained. As a result, there was a need for further RV observations at a higher observational cadence. We therefore began an intensive observing campaign with the Tillinghast Reflector Echelle Spectrograph (TRES) on the 1.5~m telescope at the Fred Lawrence Whipple Observatory, Mt. Hopkins, AZ, USA. In the 2024-5 observing season we collected 29 TRES spectra of \thisstar{}, with a median interval between successive observations of 6~days. The higher cadence of these observations allowed for a strong detection of the planetary orbit, despite the comparatively low reflex semi-amplitude. During the final stages of data collection we acquired three further TRES spectra in early 2026, immediately before the star set for the season, resulting in a final set of 32 TRES epochs.

To determine the RVs, we perform spectral extraction and multi-order cross correlations following the process outlined in \citet{Buchhave2010}. We account for variations in the instrument zero-point through nightly observations of standard stars as in \citet{Zhou2023}. The extracted RVs achieve a median precision of 15~\ms{}. Though the RV precision of the TRES observations is significantly lower than achieved with PFS, the dense sampling across three orbital revolutions of \thisstarb{} in the 2024-5 season provided improved constraints on the orbital eccentricity. Though the relative importance of the TRES data decreased as we collected more PFS RVs in 2025, this still significantly influences the final model. We present the TRES RVs collected for this work in Appendix~\ref{appendix:RV}.

\subsection{Speckle Imaging} \label{subsec:imaging}

We obtained high-resolution speckle imaging observations of \thisstar{} in order to check for close-separation stellar companions that could confound the planetary interpretation of the TESS planet candidate.
We used the Zorro speckle instrument at the Gemini South Telescope \citep{Scott2021, Howell2025} to obtain high-contrast speckle imaging data on 2024-07-14. Observations were taken in two medium-band filters with central wavelengths of 562~nm and 832~nm respectively, with a total field-of-view of 2.8$\times$2.8". We reduced the imaging data following methods described in \citet{Howell2011, Howell2016}.
We reproduce the reconstructed image and contrast limits in Appendix~\ref{appendix:speckle}. In both filters observations achieve a contrast limit of $\Delta\approx4-5$~mag at 0.1", improving to $\Delta\approx5.5$~mag by 1". We do not detect any additional stars in the speckle imaging field.

\section{Analysis} \label{sec:analysis}

\subsection{Stellar Parameters} \label{subsec:star}

\begin{deluxetable*}{lrr}[ht!]
\label{tab:star}
\centering
\tablecaption{Stellar properties of \thisstar{}.}
\tablehead{\colhead{Parameter} & \colhead{Value} & \colhead{Reference}}
\startdata
~~Right Ascension $\alpha_{\text{J2000}}$ \dotfill & 00:11:34.22 & \citet{GaiaDR3} \\
~~Declination $\delta_{\text{J2000}}$ \dotfill & +04:46:01.64 & \citet{GaiaDR3} \\
~~$V$ (mag) \dotfill & $8.70\pm0.02$ & \citet{Tycho2} \\
~~$T$ (mag) \dotfill & $8.189\pm0.006$ & \citet{Stassun2019} \\
~~Parallax $\varpi$ (mas) \dotfill & $6.008\pm0.019$ & \citet{GaiaDR3} \\
~~Distance (pc) \dotfill & $165.4\pm0.6$ & \citet{BailerJones2021} \\
~~R.A. proper motion $\mu_\alpha$ (\masyr{}) \dotfill & $+35.578\pm0.024$ & \citet{GaiaDR3} \\
~~Declination proper motion $\mu_\delta$ (\masyr{}) \dotfill & $+21.080\pm0.019$ & \citet{GaiaDR3} \\
~~Radial velocity (km~s$^{-1}$) \dotfill & $11.79\pm0.02$ & This work \\
~~$U$ (km~s$^{-1}$) \dotfill & $-34.09\pm0.10$ & This work \\
~~$V$ (km~s$^{-1}$) \dotfill & $+4.82\pm0.02$ & This work \\
~~$W$ (km~s$^{-1}$) \dotfill & $-5.38\pm0.02$ & This work \\
\hline
~~$B_\text{T}$ (mag) \dotfill & $9.419\pm0.021$ & \citet{Tycho2} \\
~~$V_\text{T}$ (mag) \dotfill & $8.771\pm0.016$ & \citet{Tycho2} \\
~~$G$ (mag) \dotfill & $8.589\pm0.02$ & \citet{GaiaDR3} \\
~~$G_{\text{BP}}$ (mag) \dotfill & $8.871\pm0.02$ & \citet{GaiaDR3} \\
~~$G_{\text{RP}}$ (mag) \dotfill & $8.140\pm0.02$ & \citet{GaiaDR3} \\
~~$J$ (mag) \dotfill & $7.659\pm0.026$ & \citet{2MASS} \\
~~$H$ (mag) \dotfill & $7.515\pm0.046$ & \citet{2MASS} \\
~~$K_S$ (mag) \dotfill & $7.349\pm0.020$ & \citet{2MASS} \\
~~$W1$ (mag) \dotfill & $7.214\pm0.035$ & \citet{AllWISE} \\
~~$W2$ (mag) \dotfill & $7.345\pm0.021$ & \citet{AllWISE} \\
~~$W3$ (mag) \dotfill & $7.346\pm0.018$ & \citet{AllWISE} \\
~~$W4$ (mag) \dotfill & $7.103\pm0.112$ & \citet{AllWISE} \\
\hline
~~Spectral type \dotfill & F8IV & \citet{Cannon1918}; This work \\
~~$T_{\text{eff}}$ (K) \dotfill & $6070\pm50$ & This work \\
~~[M/H] (dex) \dotfill & $+0.09\pm0.08$ & This work \\
~~Spec. log~$g$ ($\log$ cm~s$^{-2}$) \dotfill & $3.86\pm0.10$ & This work \\
~~$v_\text{broad}$ (k\ms{}) * \dotfill & $10.0\pm0.5$ & This work \\
~~Interstellar extinction $A_\text{v}$ (mag) \dotfill & $0.09\pm0.04$ & This work \\
~~$M_*$ ($M_\odot$) \dotfill & $1.60\pm0.04$ & This work \\
~~$R_*$ ($R_\odot$) \dotfill & $2.50^{+0.07}_{-0.06}$ & This work \\
~~$\rho_*$ (g~cm$^{-3}$) \dotfill & $0.145^{+0.012}_{-0.011}$ & This work \\
~~Model log~$g$ ($\log$ cm~s$^{-2}$) \dotfill & $3.846\pm0.024$ & This work \\
~~$L_*$ ($L_\odot$) \dotfill & $7.6^{+0.5}_{-0.4}$ & This work \\
~~Age (Gyr) ** \dotfill & $2.1\pm0.3$ & This work \\
\enddata
\tablecomments{(*) Includes both $v\sin i$ and macroturbulence. (**) Incorporates systematic age uncertainty (see text).}
\end{deluxetable*}

The nature of \thisstar{} has escaped detailed scrutiny in previous literature. In the Henry Draper catalogue, the star was given a spectral type of F8 \citep{Cannon1918}. A modern perspective on the physical parameters of the star can be found in the TESS Input Catalog \citep[TIC;][]{Stassun2018, Stassun2019}; there \thisstar{} has effective temperature $T_{\text{eff}}=6140\pm120$~K, surface gravity $\log g=3.71\pm0.08$~$\log$(\cmssq{}), radius $R_*=2.51\pm0.11~R_\odot$, and mass $M_*=1.16\pm0.16~M_\odot$. The effective temperature appears reasonably similar to the Sun, but the low surface gravity and large radius suggest that \thisstar{} is evolved. As stellar masses in the TIC are derived from a $T_{\text{eff}}$ -- mass scaling relation calibrated to the main sequence \citep{Stassun2018, Stassun2019}, these values for the mass and $\log g$ may be underestimated.

In order to precisely characterise its planet, we endeavour to further constrain the parameters of the star. To extract information on the parameters that can be derived from the TRES spectra, we use the Stellar Parameter Classification tool \citep[SPC;][]{Buchhave2012} as applied in \citet{Bieryla2024}. From this, we estimate $T_{\text{eff}}=6070\pm50$~K, $[\text{M/H}]=+0.09\pm0.08$~dex, $\log g=3.86\pm0.10$~$\log$(\cmssq{}), and a line broadening of $10.0\pm0.5$~k\ms{}. The effective temperature accords with the historical F8 spectral type, but the low surface gravity clearly indicates an evolved star. 

Next we model the physical parameters of \thisstar{} using the MIST isochrones \citep{Dotter2016, Choi2016}. Here we use the \texttt{isochrones} package \citep{isochrones} for interpolating the MIST isochrones and modelling the interstellar extinction. As input data for this model, we use space-based photometry from Tycho-2, \textit{Gaia}, 2MASS, and WISE, and apply priors on the effective temperature and metallicity from the SPC results. Given the significant distance to \thisstar{} \citep[$165.4\pm0.6$~pc,][]{BailerJones2021}, we must account for the impact of interstellar extinction on the photometry; however, we find that the combination of the input photometry and the spectroscopic prior on the effective temperature provides a strong constraint on extinction with no further information.

Using this model we estimate a stellar mass of $M_*=1.60\pm0.04~M_\odot$, radius $R_*=2.50^{+0.07}_{-0.06}~R_\odot$, and surface gravity of $\log g=3.846\pm0.024$~$\log$(\cmssq{}). The mass is notably higher than the TIC estimate, which is explicable given that the latter is based on a main sequence scaling relation \citep{Stassun2018, Stassun2019}, but the continued finding of a large radius and low surface gravity confirm that the star is substantially evolved. Given the $6070\pm50$~K effective temperature and a luminosity of $L_*=7.6^{+0.5}_{-0.4}~L_\odot$, our interpretation of the evolutionary position of \thisstar{} is that it lies at the base of the subgiant branch. We therefore designate the star as luminosity class~IV. We present the stellar parameters from our isochrone model in Table~\ref{tab:star}.

Since the star is evolved, our isochrone model provides a relatively precise stellar age constraint of $2.1\pm0.2$~Gyr. However, this estimate neglects systematic uncertainties in stellar evolutionary models, which are likely to contribute significantly to the error budget at this level of precision. To quantify this we use the \texttt{kiauhoku} package \citep{Tayar2022}, which compares the differences in physical parameters found given the same observational inputs between four sets of stellar models (MIST, YREC, Dartmouth, and GARSTEC). For the properties of \thisstar{} \texttt{kiauhoku} returns uncertainty terms in mass and age of $\pm0.015~M_\odot$ and $\pm0.2$~Gyr respectively. While the error in mass is small compared to the existing statistical uncertainty, the systematic error in mass is equivalent to the statistical uncertainty. Combining the two values in quadrature, we find an increased age uncertainty of $\pm0.3$~Gyr. We report this as our adopted age in Table~\ref{tab:star}.

\subsubsection{Asteroseismic Mass}

Historically, mass estimation for planet-hosting evolved stars has been surrounded by controversy. The root issue is whether the stellar masses of evolved stars with isochronal mass estimates above $M_*\gtrsim1.5~M_\odot$ are systematically overestimated (e.g. \citealt{Lloyd2011, Johnson2013}; see the introduction of \citealt{Ghezzi2018} for an overview). Our isochrone mass for \thisstar{}, $M_*=1.60\pm0.04~M_\odot$, places it in the region of concern, and it is notable that this significantly exceeds the mass estimate in the TIC \citep[$1.16\pm0.16~M_\odot$;][]{Stassun2019}. It would therefore be beneficial to externally validate this mass estimate.

Asteroseismology provides a powerful method for mass estimation of stars with solar-like oscillations, including evolved stars \citep{Chaplin+Miglio2013}. Asteroseismic detections in evolved planet hosts observed by \textit{Kepler}, K2, and TESS have made significant contributions understanding their mass distribution \citep[e.g.][]{North2017, Malla2020, Malla2024}, so we investigate whether solar-like oscillations of \thisstar{} can be detected in the TESS photometry. Given only three sectors of TESS data, detecting oscillations in a $T=8.2$ subgiant is challenging, but not impossible \citep[e.g.][]{Zhou2024}.

The power spectrum of the TESS light curve of \thisstar{} shows a weak hump of excess power near 800~$\mu$Hz that is consistent with solar-like oscillations.  Using the \texttt{pySYD} pipeline \citep{Huber2009, Chontos2022} we measure a value of $\nu_{\rm max} = 803 \pm 47~\mu$Hz.  However, a reliable value for the so-called large frequency separation ($\Delta\nu$) could not be estimated from the available TESS photometry.

To calculate the mass of \thisstar{} we use the following scaling relation \citep{Brown1991, Kjeldsen+Bedding1995, Stello2008}:
\begin{equation}
    \frac{\nu_{\rm max}}{\nu_{\rm max,\odot}} \approx \frac{(M/M_\odot)(T_{\rm eff}/5772~{\rm K})^{3.5}}{L/L_\odot}
    \:,
\end{equation}
where $\nu_{\rm max,\odot} = 3090 \pm 30~\mu$Hz \citep{Huber2011}. Using the effective temperature and luminosity from Table~\ref{tab:star}, this results in an asterosesimic mass of $1.66 \pm 0.15~M_\odot$. This shows excellent agreement with the isochrone mass estimate, and seemingly confirms the relatively high mass of \thisstar{}.


\subsubsection{Kinematic Age}

An independent check on the age of \thisstar{} can be derived from its kinematics. When stars form, they begin with space velocities close to the mean velocity of the galactic disk. However, over time gravitational stirring within the galactic disk gradually increases the velocity dispersion \citep{AlmeidaFernandes2018}. Hence the space velocity of an individual star encodes information on its total age, where larger space velocities imply an older age. We calculate the Sun-relative relative space velocities of \thisstar{} following \citet{Johnson1987}, using the position, proper motion, and radial velocity from \textit{Gaia}~DR3. For the conversion between equatorial and galactic coordinates, we use the \textit{Hipparcos} transformation matrix, also utilised in \textit{Gaia}~DR3.\footnote{See Section~4.1.7 (Equation 4.62) of the \textit{Gaia}~DR3 documentation \\(\url{https://www.cosmos.esa.int/web/gaia-users/archive/gdr3-documentation}).} In this way we find $(U, V, W)=(-34.09\pm0.10, +4.82\pm0.02, -5.38\pm0.02)$~k\ms{}.

To quantify the kinematic age of \thisstar{}, we use the $UVW$-age relationships of \citet{AlmeidaFernandes2018} and \citet{Veyette2018}. These relationships are calibrated to Sun-like thin disk stars in the solar neighbourhood with known isochrone ages, differing in the details of calibrator selection in that \citet{Veyette2018} restricts the sample to lower-mass stars to reduce a bias towards younger ages. Using the \citet{Veyette2018} relation we estimate a kinematic age of $2.3^{+3.0}_{-1.3}$~Gyr for \thisstar{}. Though the uncertainties on this age estimate are much greater than for the isochrone age ($2.1\pm0.3$~Gyr), the two appear to be in good agreement. This broadly supports the accuracy of our adopted stellar parameters.

\subsection{Identification of a Stellar Companion}

The speckle imaging observations recorded in Section~\ref{subsec:imaging} did not detect any stellar companions within about $<$1.2" of \thisstar{}; however, stellar companions at wider separations would have been missed in this data. To expand our search radius we turn to \textit{Gaia}, which is highly efficient for the detection of binaries with projected separations beyond $\sim$1". For this purpose we query the \textit{Gaia}~EDR3 binary catalogue constructed in \citet{ElBadry2021}. In this catalogue, \thisstar{} (source~ID 2741377642494388864) is identified to have a stellar companion (source~ID 2741377638198866304) at a projected separation of 4.12", equivalent to 685~AU at the distance of \thisstar{}, and a position angle of 233.8$\degree$. This star has a parallax ($\varpi=6.039\pm0.075$~mas) and proper motion ($\mu_\alpha,\mu_\delta=[+36.227\pm0.089,+22.203\pm0.070]$~\masyr{}) that agree closely with \thisstar{}, such that in the \citet{ElBadry2021} catalogue the chance alignment probability for the two stars is essentially negligible ($\approx6\times10^{-7}$). We therefore designate this star as \thisstar{}~B.

In \textit{Gaia}~DR3, \thisstar{}~B has a \textit{G}-band magnitude of 15.2 but no colour \textit{BP}/\textit{RP} photometry, presumably due to the deleterious impact of the much brighter \thisstar{}~A on its \textit{Gaia} data. As a result we may estimate an absolute magnitude of $M_G=9.1$, but without colour information the physical nature of \thisstar{}~B is uncertain. If it is a main sequence star, it is likely to be an early-M dwarf; a nearby star of similar absolute magnitude is Gliese~887 ($M_G=8.9$), which has a spectral type of M2V. However, it could also conceivably be an unresolved M-dwarf binary, or relatively bright white~dwarf.

Given the unknown nature of \thisstar{}~B, we also cannot precisely quantify its impact on the TESS observations. Flux contamination from stellar companions and background stars can have a significant effect on radius estimation for transiting planets \citep[e.g.][]{Furlan2017, Ziegler2020, Han2025}. However, in the \textit{Gaia} \textit{G}-band, the 6.6-mag contrast between \thisstar{}~A and B implies a flux ratio of only 0.2\%. This cannot be exactly mapped onto the contrast in the TESS band since the TESS bandpass extends further redward, but regardless of the properties of \thisstar{}~B its impact on the TESS photometry is likely to be relatively small, so we do not attempt to correct for this in this work.

To check for additional more widely-separated stellar companions to \thisstar{}, we query \textit{Gaia}~DR3 with a large search radius of 2000", equivalent to a projected separation of ~1.6~pc at the distance of the system. Other than \thisstar{}~B, we do not detect any additional stars with parallax and proper motion consistent with \thisstar{} in this search radius. We therefore conclude that \thisstar{} is a binary, and has no further stellar companions.

\subsection{Transit and RV Model}

\begin{figure*}
    \centering
    \includegraphics[width=\textwidth]{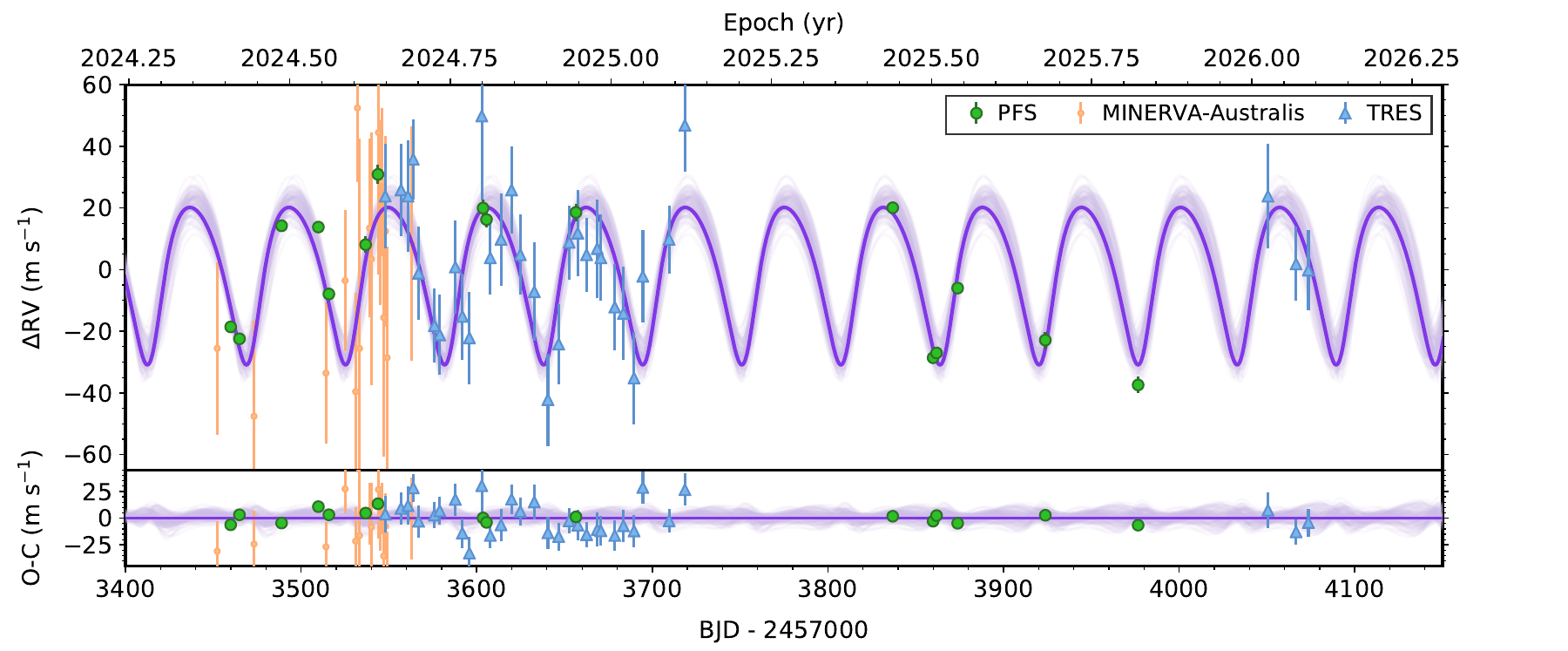}
    \includegraphics[width=\textwidth, clip=true, trim=0cm 0cm 0cm 1cm]{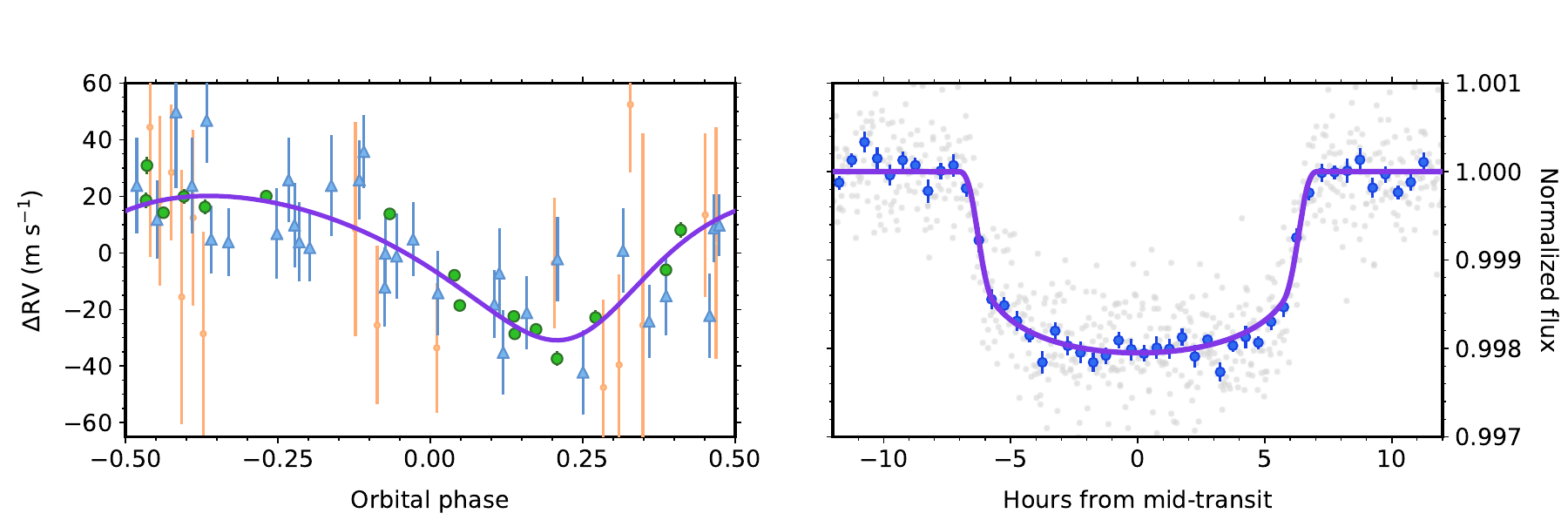}
    \caption{Our joint transit + RV model for \thisstarb{}. (\textit{Top}) RV data and model. The dark model line corresponds to the best-fit solution, whereas the light model lines are drawn randomly from the posteriors.
    (\textit{Lower left}) RV data and best-fit model phased to the orbital period. (\textit{Lower right}) The single observed TESS transit with best-fit model. The light points are the individual 2-min cadence flux measurements, while the dark points are 30-min bins.
    }
    \label{fig:model}
\end{figure*}

Visual inspection of our RV data reveals significant variation which appears to be periodic. To quantify the significance of this, we calculate a Lomb-Scargle periodogram for the RVs using \texttt{astropy.timeseries.LombScargle}. We identify a single strong peak at $P\approx56$~days. This corresponds almost exactly to the median value for the orbital period estimated from the transit properties, 
$55^{+28}_{-17}$~d (Section~\ref{subsec:TESS}), and appears fully consistent with the RV signal expected from the transiting planet candidate. We therefore move directly to jointly fitting the TESS transit with our RVs.

The model used here for fitting the RV data has been adapted from its implementation in \citet{Venner2021}, while we use the \texttt{batman} package \citep{batman} to generate models for the transit. As input data we incorporate all of our RV data and the TESS photometry surrounding the single observed transit in Sector~70.

Our joint model involves a total of 18 variable parameters, which can be described as follows. To represent the star we use the stellar mass $M_*$ and density $\rho_*$. Four parameters describe the orbital properties of the planet, being the orbital period $P$, the eccentricity $e$ and argument of periastron $\omega$ (parameterised as $\sqrt{e}\sin\omega$ and $\sqrt{e}\cos\omega$), and the time of inferior conjunction $T_0$. One further variable is required to model the RVs, the semi-amplitude parameter $K$, whereas four are required for the transit, being the impact parameter $b$, the planet-star radius ratio $R_p/R_*$, and the quadratic limb darkening coefficients $u_1$ and $u_2$. Finally, there are six variables to describe the RV offsets $\gamma$ and RV jitters $\sigma$ for the three RV datasets, and one parameter to parameterise the dispersion in the TESS light curve $\sigma_\text{TESS}$.

All priors used in the model are listed in Table~\ref{tab:planet}. For the stellar mass and density we use priors derived from the values in Table~\ref{tab:star} sampled in logarithm (i.e. $\log M_*=0.203\pm0.011~\log M_\odot$; $\log \rho_*=-0.840\pm0.034~\log\text{g~cm}^{-3}$). For the orbital period $P$, we apply a uniform box prior of [55, 58]~d approximately centred on the period peak identified in the Lomb-Scargle periodogram. We sample the RV semi-amplitude $K$, radius ratio $R_p/R_*$, and jitter parameters log-uniformly. For the limb darkening coefficients, we apply Gaussian priors derived from the \citet{Claret2018} models in the TESS bandpass (i.e. $u_1=0.35\pm0.07$, $u_2=0.21\pm0.07$).

We use the Markov Chain Monte Carlo ensemble sampler \texttt{emcee} \citep{emcee} to explore the parameter space of our model. A total of 75 walkers were used to sample the model for $3\times10^5$ steps, with MCMC convergence evaluated by comparison to the autocorrelation timescale as in \citet{Venner2021}. We then discarded the first half of the chain as burn-in and saved every 150$^\text{th}$ step to extract our final posteriors.

\section{Results} \label{sec:results}

\begin{deluxetable*}{lrr}[ht!]
\label{tab:planet}
\centering
\tablecaption{Parameters of \thisstarb{} from our transit + RV joint model.}
\tablehead{\colhead{Parameter} & \colhead{Prior} & \colhead{Value}}
\startdata
~~Stellar mass $M_*$ ($M_\odot$) & $10^{\mathcal{N}(0.203, 0.011)}$ & $1.60\pm0.04$ \\
~~Stellar density $\rho_*$ ($\rho_*$) & $10^{\mathcal{N}(-0.840\pm0.034)}$ & $0.145^{+0.012}_{-0.011}$ \\
~~Stellar radius $R_*$ ($R_\odot$) & Derived & $2.50^{+0.07}_{-0.06}$ \\
\hline
~~Orbital period $P$ (days) * & $\mathcal{U}(55, 58)$ & $56.40^{+0.19}_{-0.23}$ \\
~~Time of inferior conjunction $T_0$ (BJD) & $\mathcal{U}(2460231.5, 2460231.6)$ & $2460231.5482\pm0.0016$ \\
~~$\sqrt{e}\sin\omega_p$ & $\mathcal{U}(-1, 1)$ & $-0.12^{+0.21}_{-0.16}$ \\
~~$\sqrt{e}\cos\omega_p$ & $\mathcal{U}(-1, 1)$ & $-0.41^{+0.14}_{-0.09}$ \\
~~RV semi-amplitude $K$ (\ms{}) & $\mathcal{U}(1/K)$ & $25.3\pm2.2$ \\
~~Impact parameter $b$ & $\mathcal{U}(0, 1+R_p/R_*)$ & $0.56^{+0.09}_{-0.15}$ \\
~~Radius ratio $R_p/R_*$ & $\mathcal{U}(R_*/R_p)$ & $0.0430^{+0.0008}_{-0.0009}$ \\
~~Linear limb darkening $u_1$ & $\mathcal{N}(0.35, 0.07)$ & $0.35^{+0.06}_{-0.05}$ \\
~~Quadratic limb darkening $u_2$ & $\mathcal{N}(0.21, 0.07)$ & $0.23^{+0.07}_{-0.06}$ \\
\hline
~~Semi-major axis $a$ (AU) & Derived & $0.336\pm0.003$ \\
~~Normalised semi-major axis $a/R_*$ & Derived & $28.9^{+0.8}_{-0.7}$ \\
~~Eccentricity $e$ & Derived & $0.21^{+0.07}_{-0.08}$ \\
~~Argument of periastron $\omega_p$ (degrees) & Derived & $196\pm27$ \\
~~Time of periastron $T_\text{peri}$ (BJD) & Derived & $2460244.0^{+4.9}_{-4.4}$ \\
~~Orbital inclination $i$ (degrees) & Derived & $88.91^{+0.24}_{-0.11}$ \\
~~Transit duration $T_D$ (hours) & Derived & $13.52^{+0.16}_{-0.14}$ \\
~~Planetary mass $M_p$ ($M_J$) & Derived & $0.635\pm0.056$ \\
~~Planetary radius $R_p$ ($R_J$) & Derived & $1.07\pm0.04$ \\
~~Density $\rho_p$ (g cm$^{-3}$) & Derived & $0.69^{+0.11}_{-0.09}$ \\
~~Incident flux $I_p$ ($I_\oplus$) & Derived & $67.3^{+4.5}_{-4.0}$ \\
~~Maximum equilibrium temperature ($T_\text{eq}$) & Derived & $798^{+13}_{-12}$ \\ 
\hline
~~PFS RV offset $\gamma_\text{PFS}$ (\ms{}) & $\mathcal{U}(0, 20)$ & $+7.5\pm2.0$ \\
~~MINERVA-Australis RV offset $\gamma_\text{M-A}$ (\ms{}) & $\mathcal{U}(10980, 11030)$ & $+11007\pm9$ \\
~~TRES RV offset $\gamma_\text{TRES}$ (\ms{}) & $\mathcal{U}(-70, -30)$ & $-47.8^{+2.9}_{-2.7}$ \\
~~PFS RV jitter $\sigma_\text{PFS}$ (\ms{}) & $\mathcal{U}(1/\sigma_\text{PFS})$ & $6.4^{+2.2}_{-1.5}$ \\
~~MINERVA-Australis RV jitter $\sigma_\text{M-A}$ (\ms{}) & $\mathcal{U}(1/\sigma_\text{M-A})$ & $17^{+12}_{-11}$ \\
~~TRES RV jitter $\sigma_\text{TRES}$ (\ms{}) & $\mathcal{U}(1/\sigma_\text{TRES})$ & $5.4^{+4.4}_{-3.6}$ \\
~~TESS noise $\sigma_\text{TESS}$ (ppm) & $\mathcal{U}(1/\sigma_\text{TESS})$ & $423^{+15}_{-13}$ \\
\enddata
\tablecomments{(*) Based only on the RVs and the transit observed in TESS Sector 70. The orbital period can be further constrained by the absence of transits in Sectors 42 and 43; see Section~\ref{subsec:period_refinement}.}
\end{deluxetable*}

We show the results of our transit + RV model in Figure~\ref{fig:model}, and the corresponding parameters and 68.3\% confidence intervals in Table~\ref{tab:planet}. We identify a unique joint solution corresponding to the planet candidate detected from the TESS Sector~70 single transit, which we can therefore confirm as a planet, \thisstarb{}.

Proceeding from our orbital period initialisation we find a narrow, unimodal constraint of $P=56.40^{+0.19}_{-0.23}$~d, and an RV semi-amplitude of $K=25.3\pm2.2$~\ms{} that is more than $>$10$\sigma$ above zero. The eccentricity is reasonably well-constrained, with $e=0.21^{+0.07}_{-0.08}$ and $\omega_p=196\pm27\degree$. The comparatively large uncertainties on these parameters likely reflects the relatively low ratio between the RV semi-amplitude $K$ and the RV jitter (e.g. $\sigma_\text{PFS}=6.4^{+2.2}_{-1.5}$~\ms{}), which makes it challenging to detect the relatively subtle effect of eccentricity on the RV curve. The eccentricity can be further constrained in part by the transit, but this is limited by the impact parameter which is likewise somewhat uncertain ($b=0.56^{+0.09}_{-0.15}$). Regarding its physical properties, \thisstarb{} has a mass of $M_p=0.635\pm0.056~M_J$ and radius of $R_p=1.07\pm0.04~R_J$, corresponding to a density of $\rho_p=0.69^{+0.11}_{-0.09}$~g~cm$^{-3}$. We estimate a semi-major axis of $a=0.336\pm0.003$~AU, which entails that the planet receives an average incident flux of
$I_p=67.3^{+4.5}_{-4.0}~I_\oplus$. The corresponding maximum value for the equilibrium temperature, given zero albedo, is therefore $T_\text{eq}=798^{+13}_{-12}$~K. The sum of all of these properties confirm that \thisstarb{} is a \textit{bona fide} Warm Jupiter.

The standard deviation of the RV residuals for the PFS, TRES, and MINERVA-Australis datasets are 5.5~\ms{}, 16~\ms{}, and 28~\ms{} respectively. For the latter two datasets, this almost exactly matches the median uncertainties of 15~\ms{} and 31~\ms{}, which is perhaps unsurprising on account of the comparatively large uncertainty terms. However, the residuals for the PFS data significantly exceeds the 2.1~\ms{} median uncertainty (Section~\ref{subsec:PFS}), necessitating a significant jitter term in the model ($\sigma_\text{PFS}=6.4^{+2.2}_{-1.5}$~\ms{}). The source of this excess scatter is not entirely clear; the residual variations are not identifiably periodic in the present dataset, so it is perhaps most likely that the excess RV variability is driven by stellar effects such as pulsations. Whatever its origin, the excess scatter ultimately sets a limit on the precision of the planetary parameters.

\subsection{Refining the Orbital Period Constraint} \label{subsec:period_refinement}

\begin{figure*}[t!]
    \centering
    \includegraphics[width=\textwidth]{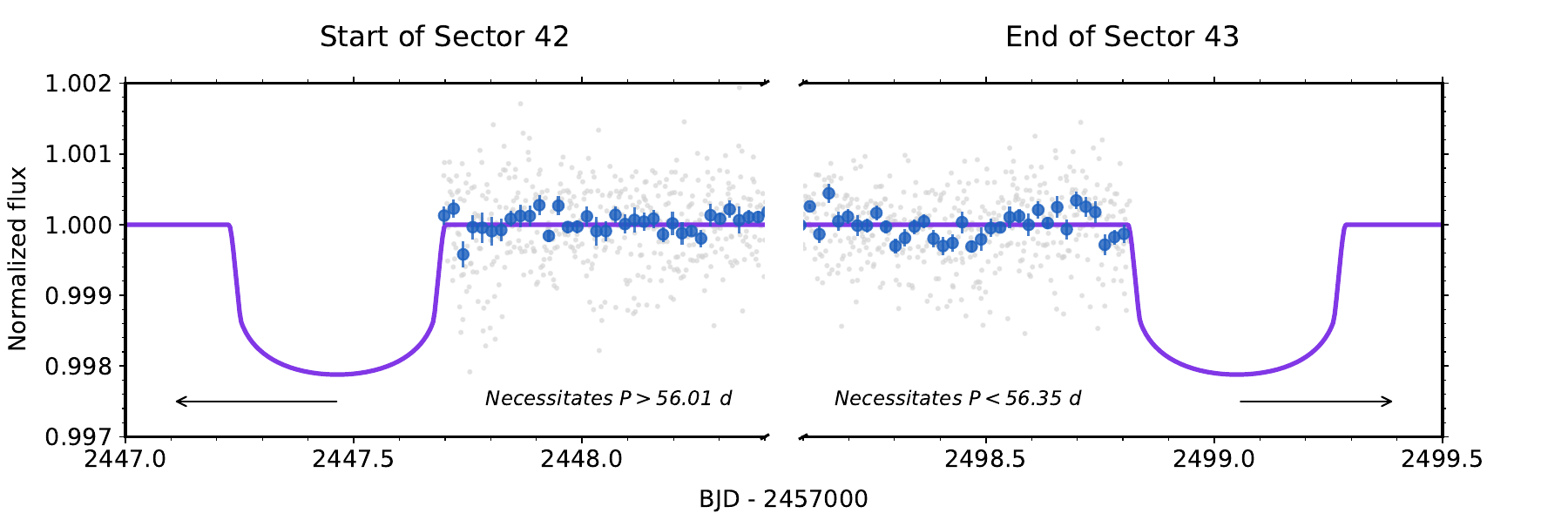}
    \includegraphics[width=\textwidth]{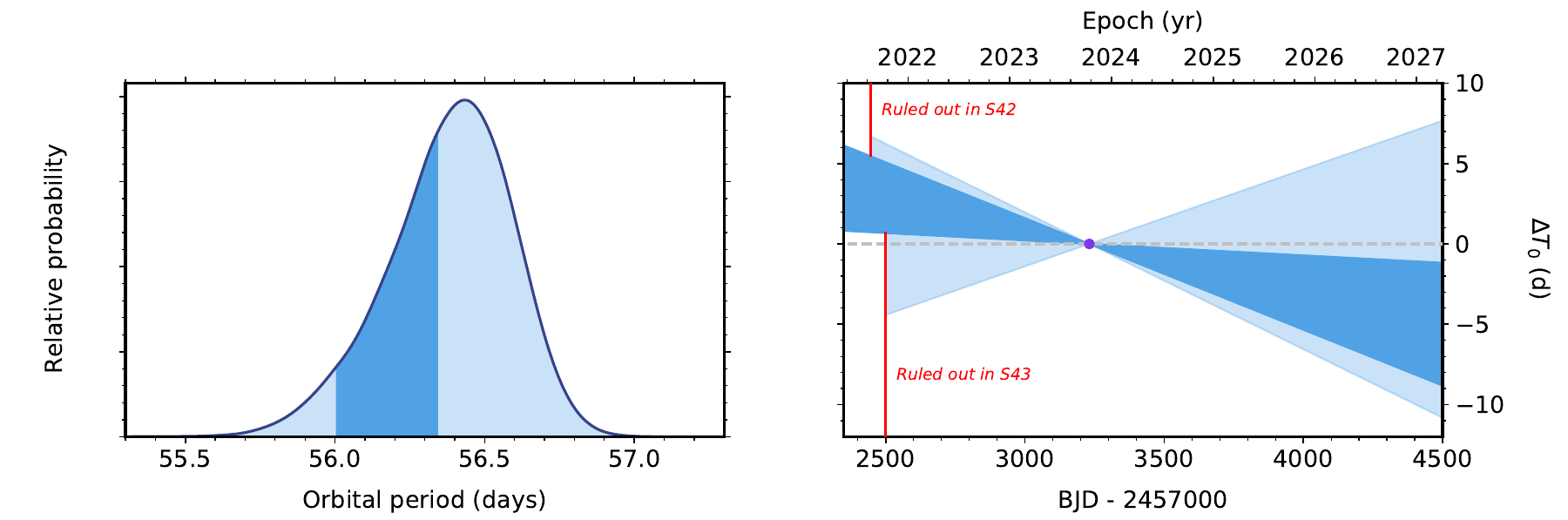}
    \caption{Refinements on the orbital period constraint from prior TESS sectors. (\textit{Top}) Inset of TESS Sector~42 and 43 observations. Light points are 2-min cadence measurements, while dark points are binned to 30 mins. We visualise the minimum and maximum orbital period for \thisstarb{} allowed by the absence of transits in these sectors by showing the best-fit model for the transit observed in Sector~70 at the respective latest and earliest transit times consistent with the data.    
    (\textit{Lower left}) Period posterior distribution from our joint model (Figure~\ref{fig:model}) showing the refined constraint afforded by the absence of transits in Sectors~42 and 43. The dark shaded region between $[56.01, 56.35]$~d reflects final range of allowed orbital periods.
    (\textit{Lower right}) Transit time prediction given our orbital period constraints. The light shaded region reflects the 2$\sigma$ confidence interval from our joint model, normalised to the median value of $56.40$~days. The red lines show the transit times that can be filtered out by the Sector~42 and 43 non-detections, and the dark shaded region shows the remaining allowed orbital periods, i.e. $[56.01, 56.35]$~d (note that this actually excludes the original median period). At 2026.5, we find that the refined range on the transit epoch is 6~days.
    }
    \label{fig:backtrack}
\end{figure*}

Our RV follow-up has revealed the orbital period of \thisstarb{}, which raises the question of the feasibility of observing a second transit. The orbital period from our joint model, $56.40^{+0.19}_{-0.23}$~d, corresponds to a fractional uncertainty of $\sim$0.4\%. Nominally this is quite a high level of precision, but it is nonetheless insufficient to predict the transit ephemeris with the precision necessary to easily re-observe the transit; we calculate that by 2026.5, when the star will next be observable, the 1$\sigma$ uncertainty on the transit time will be slightly over 7~days and the 2$\sigma$ uncertainty nearly 15~days, more than an order of magnitude larger than the transit duration. Thus, given our present constraint on the orbital period, recovering the transit of \thisstarb{} would be a significant undertaking. While further RV observations would lead to smaller uncertainties, the improvement is likely to be small compared to the total transit time uncertainty.

However, in the case of this system, we may significantly refine the constraint on the orbital period by utilising the whole set of TESS photometry. As described in Section~\ref{subsec:TESS}, TESS has observed \thisstar{} in a total of three sectors; the last, Sector~70, contains a transit of \thisstarb{}, whereas the other two, Sectors~42 and 43, do not. Projecting our best-fit solution backwards, we estimate that the transit that occurred 14 revolutions prior to Sector~70 occurred at TBJD~$2442.0^{+3.2}_{-2.7}$, whereas the subsequent transit occurred at TBJD~$2498.4^{+3.0}_{-2.5}$ (1$\sigma$ uncertainties). These epochs happen to correspond closely to the beginning of Sector~42 data and the end of Sector~43 data at TBJD~$[2447.7, 2498.8]$ respectively. Visual inspection of the TESS photometry reveals no evidence for transits of \thisstarb{} in either sector, even in mid-sector observations that were strongly affected by scattered light (Figure~\ref{fig:lightcurve}). It therefore appears that TESS narrowly missed two transits of \thisstarb{} before and after the respective end of science observations in these two sectors.

In the upper panel of Figure~\ref{fig:backtrack} we visualise the constraints on the orbital period of \thisstarb{} provided by non-detections in the Sector~42 and 43. We show our best-fit model for the observed Sector~70 transit translocated onto the respective latest and earliest times of transit allowed by the existing data. Given the known transit depth and duration for \thisstarb{}, we find that the mid-time of the transit before Sector~42 must have occurred prior to TBJD~$<$~2447.47, whereas the transit after Sector~43 must have occurred subsequent to TBJD~$>$~2499.05. This necessitates that $P$ must be between $[56.01, 56.35]$~d, otherwise a transit would have occurred in one or both sectors. Notably, the constraint from Sector~43 actually rules out the $56.40$~d median period from our joint model.

In the lower panels of Figure~\ref{fig:backtrack} we show the impact of this refined period constraint on the posterior from our joint model and on the prediction for the transit time of \thisstarb{}. Whereas the original 2$\sigma$ uncertainty in transit epoch at 2026.5 spans $\sim$15~days, incorporating the restrictions from Sector~42 and 43 results in a refined \textit{maximum} range in transit epoch of 6~days, sharply reducing the uncertainty on the transit time prediction. Incorporating the constraints provided by the preceding TESS sectors effectively removes 65\% of the initially allowed period posterior space.

\section{Discussion} \label{sec:discussion}

In many respects \thisstarb{} appears to be a typical Warm Jupiter, with unremarkable mass and radius of $M_p=0.635\pm0.056~M_J$ and $R_p=1.07\pm0.04~R_J$ and a relatively low orbital eccentricity of $e=0.21^{+0.07}_{-0.08}$. The orbital period is comparatively long for a transiting Warm Jupiter, $P=56.40^{+0.19}_{-0.23}$~d (further refined to $[56.01, 56.35]$~d due to the absence of further transits in previous TESS sectors). There are comparatively few transiting Warm Jupiters known to have similar orbital periods, undoubtedly due to the challenges involved in detecting such long-period planets in transit. A unique factor is the comparatively bright host star, $V=8.70$, which has greatly facilitated RV follow-up observations and will enable further future observations; among similar TESS planet hosts this is most comparable to HD~1397 \citep[$V=7.79$,][]{Nielsen2019, Brahm2019} and HD~221416 \citep[$V=8.15$,][]{Huber2019}, both of which have significantly shorter orbital periods ($P=11.5$~d and 14.3~d respectively). Furthermore, the stellar mass of $M_*=1.60\pm0.04~M_\odot$ is among the largest of any currently known Warm Jupiter host.

Observationally, \thisstarb{} is perhaps most similar to NGTS-20~b, a $1.07\pm0.04~R_J$ Warm Jupiter that orbits a moderately evolved $1.47\pm0.09~M_\odot$ star with an orbital period of $54.19$~d \citep{UlmerMoll2022}. However, NGTS-20~b has a higher mass ($2.98^{+0.16}_{-0.15}~M_J$) and orbital eccentricity ($0.432\pm0.023$) than \thisstarb{}. Another Warm Jupiter with similar properties is the recently confirmed KOI-134~b, which orbits a $1.407\pm0.043~M_\odot$ star located near the main sequence turn-off with period $67.1277^{+0.0045}_{-0.0057}$~d and eccentricity $0.16^{+0.02}_{-0.03}$, and has mass and radius of $1.09^{+0.12}_{-0.08}~M_J$ and $1.074\pm0.017~R_J$ respectively \citep{Nabbie2025}. While these observational properties are reasonably similar to \thisstarb{}, KOI-134~b is joined by a second non-transiting sub-Saturn-mass planet, KOI-134~c, found in the interior 2:1 orbital resonance at an orbital period of $33.950^{+0.013}_{-0.020}$~d. We have inspected the RV residuals from our best-fit model, and do not observe any significant evidence for additional planets orbiting interior to \thisstarb{}. This suggests that the \thisstar{} and KOI-134 systems have different histories, despite similarities between the transiting Warm Jupiters in both systems.

\subsection{Interpreting the Orbital Eccentricity}

Following their discovery of a Warm Jupiter orbiting the subgiant star TOI-4582, \citet{Grunblatt2023} brought attention to the fact that the known Warm Jupiters around evolved stars tend to have more eccentric orbits than their cousins orbiting main sequence stars. After establishing that this could not be explained by differences in stellar mass or metallicity or due to observational bias in favour of high eccentricities, the authors proposed that this could be explained as a relationship between orbital period and eccentricity, which they quantified via a linear regression as $e=0.38\times(\log P)-0.18$ with a standard deviation of $0.05$. They find that this differs from the broader exoplanet population at $>$5$\sigma$ confidence, indicating that it reflects a marked feature of the sample of Warm Jupiters orbiting evolved stars, and suggest that the elevated eccentricities may result from planet-planet scattering. 

\thisstarb{} appears to fall short of this period-eccentricity relation. Given $P$ = 56 days the \citet{Grunblatt2023} relation ``predicts" an eccentricity of $\sim$0.49, which is about 4$\sigma$ higher than the observed value of $0.21^{+0.07}_{-0.08}$. The eccentricity of \thisstarb{} differs markedly from the three planets with orbital periods above $>$30~d in the \citet{Grunblatt2023} sample, TOI-4582~b ($P=31.03$~d, $e=0.51\pm0.05$), Kepler-432~b ($P=52.50$~d, $e=0.51\pm0.03$; \citealt{Quinn2015}), and NGTS-20~b ($P=54.19$~d, $e=0.432\pm0.023$; \citealt{UlmerMoll2022}). Thus \thisstarb{} appears to belong to a different region of eccentricity space than similar planets.

There are several reasons to be cautious in interpretation of the eccentricity of \thisstarb{}. For one, the sample of Warm Jupiters orbiting evolved stars remains quite small (the \citealt{Grunblatt2023} sample contains only 7 planets with $P>10$~d), so there is a risk of over-interpreting a limited dataset; further additions to the sample of Warm Jupiters orbiting evolved stars would be very welcome. Secondly, though the star is meaningfully evolved, \thisstar{} has different properties from many of the previously known Warm Jupiter hosts in the \citet{Grunblatt2023} sample; for example, at $M=1.60\pm0.04~M_\odot$ it is comparatively massive (equal to K2-99, the most massive member of the \citealt{Grunblatt2023} sample). The stellar mass may be relevant if the eccentricities of Warm Jupiters orbiting evolved stars are indeed excited to planet-planet scattering, as this requires the formation of systems with multiple giant planets in the appropriate range of orbital separations. Unfortunately, the occurrence statistics of Warm Jupiters as a function of stellar mass are not well-known; the nearest analogues are the occurrence rates from giant star RV surveys, which suggest a peak in occurrence around $\sim$1.7~$M_\odot$ for giant planets with $80<P<3600$~d, and the occurrence rates of Hot Jupiters orbiting main sequence stars from TESS, which suggest if anything a slight decline towards super-solar masses \citep{Zhou2019, Beleznay2022}. Since eccentricity excitation via planet-planet scattering as suggested by \citet{Grunblatt2023} requires the formation of multi-planet Warm Jupiter systems, if systems with multiple Warm Jupiters are less abundant around more massive stars then there may be no tension with the \citet{Grunblatt2023} results. \thisstar{} is also comparatively new to the subgiant branch, with $\log g=3.846\pm0.024~\log$(\cmssq{}) and $T_\text{eff}=6070\pm50$~K that lie outside of the population cuts employed by \citet{Grunblatt2023} ($\log g<3.8~\log$(\cmssq{}), $T_\text{eff}<6000$~K). Thus if orbital eccentricities are excited later in stellar evolution, we would further not expect to see an elevated eccentricity in \thisstarb{}.

At this point it appears useful to make comparison to the recent results of \citet{Fairnington2026}, who used the ``photoeccentric effect" \citep{Dawson2012} to study the distribution of warm ($8<P<200$~d) single planets detected by TESS. Focusing on their results for Jovian-sized planet candidates ($8<R_p<16~R_\oplus$), which builds on previous work from \citet{Dong2021.catalogue}, \citet{Fairnington2026} find that the eccentricity distribution for Warm Jupiters is significantly bimodal, with $41\pm13\%$ belonging to a lower-eccentricity population ($\bar{e}=0.225\pm0.050$) and a $59\pm13\%$ majority belonging to a higher-eccentricity population ($\bar{e}=0.704^{+0.062}_{-0.044}$). It is notable that the orbital eccentricity of \thisstarb{} is in good agreement with the former population, whereas the aforementioned comparanda of TOI-4582~b, Kepler-432~b, and NGTS-20~b hew closer to the latter group.

On account of the differing stellar properties of \citet{Fairnington2026} sample, it cannot be assumed that the population-level eccentricity distributions would also apply for the evolved stars at the centre of our discussion. However, it appears conceivable that the eccentricities of Warm Jupiters orbiting evolved stars could be significantly bimodal. If this is the case, further discoveries may reveal more \thisstarb{}-like Warm Jupiters with moderate eccentricities orbiting evolved stars. Alternatively, if Warm Jupiter orbital eccentricities are excited via planet-planet scattering beyond the main sequence as suggested by \citet{Grunblatt2023}, then the discovery of further planets similar to \thisstarb{} may allow for constraints on the timing of these events. The main conclusion to be drawn from this discussion, then, is that there is a critical need for the discovery and characterisation of additional Warm Jupiters orbiting massive and/or evolved stars.

\subsection{Prospects for Transit Recovery}

Much of the prospects for further study of \thisstarb{} rest on the successful detection of a second transit. Though we have achieved a comparatively precise constraint on the orbital period from RVs ($P=56.40^{+0.19}_{-0.23}$~d), and have been able to further refine the period constraint on account of the lack of additional transits in preceding TESS observations ($[56.01, 56.35]$~d), the uncertainty on the transit time remains large; by epoch 2026.5 the total range of possible transit times spans 6~days. As mentioned previously in Section~\ref{subsec:TESS}, TESS is not scheduled to observe \thisstar{} again for the foreseeable future, so observation of a second transit will depend on targeted follow-up.

Due to the long transit duration, $13.52^{+0.16}_{-0.14}$~hr, it will not be possible to capture a whole transit of \thisstarb{} from most ground-based sites. Nonetheless, similar transit detections have been achieved from the ground on the basis of partial transits. In particular, NGTS \citep{Wheatley2018} has proved especially adept at ground-based detection of long-period, long-duration transits from incomplete transit events thanks to its high photometric precision \citep[e.g.][]{Gill2020, Gill2024, Bryant2021, UlmerMoll2022}. Alternatively, utilisation of multiple sites distributed across the world can reduce the difficulties in chasing long-duration transits with uncertain ephemerides; notably the LCOGT network \citep{LCOGT} has recently been used to recover the transit of the $P=131$~d giant planet TOI-6692~b \citep{Bieryla2026}. Such observations can also benefit from amateur observations, as demonstrated by successful observations of long-period planets through the Unistellar network \citep{Perrocheau2022, Peluso2023}. The relatively small 0.2\% transit depth of \thisstarb{} may make ground-based detection more challenging, but this is not beyond the range of feasibility.

Acknowledging the long transit duration, comparatively shallow transit depth, and substantial ephemeris uncertainty, space-based observations present significant advantages for recovering the transit of \thisstarb{}. In particular, this challenge appears well-suited to the capabilities of CHEOPS \citep{CHEOPS}. CHEOPS has previously been used to detect transits of long-period planets, even comparatively shallow events such as for $\nu^2$~Lupi~d \citep[][]{Delrez2021}. Here the main remaining obstacle is the large transit window.

\section{Conclusions} \label{sec:conclusions}

In this work we have reported the discovery and confirmation of \thisstarb{}, a transiting Warm Jupiter orbiting a relatively bright ($V=8.70$) subgiant star. \thisstarb{} was detected from a single transit event observed in TESS Sector~70; we then confirmed the planetary nature of the companion using follow-up RV observations with the MINERVA-Australis, PFS, and TRES spectrographs. We determine a stellar mass of $1.60\pm0.04~M_\odot$ from isochrones and $1.66\pm0.15~M_\odot$ from TESS asteroseismology; the comparatively large mass makes this one of the most massive Warm Jupiter hosts known.

Our joint model to the transit and RV data reveal that \thisstarb{} has an orbital period of $P=56.40^{+0.19}_{-0.23}$~days, a mass of $M_p=0.635\pm0.056~M_J$, and a radius of $R_p=1.07\pm0.04~R_J$. We find that the planet has a moderate orbital eccentricity of $e=0.21^{+0.07}_{-0.08}$, which contrasts with similar previously-known Warm Jupiters orbiting evolved stars; this suggests there is a hitherto unknown diversity in orbital eccentricities for these planets. Whether this may reflect a diversity in formation histories for Warm Jupiters orbiting massive stars or the delayed onset of eccentricity excitation beyond the main sequence will depend on future discoveries.

In addition to the orbital period constraint determined from the RVs, the absence of additional transits in preceding TESS sectors allows us to refine the range in possible periods to $[56.01,56.35]$~days, effectively removing 65\% of the period posterior. While this significantly reduces the uncertainty in the projected transit ephemeris, from an initial $~$15 days (2$\sigma$) to a maximum range of 6 days at epoch 2026.5, this remains a comparatively large window for future transit searches. Nonetheless, as TESS is not expected to re-observe \thisstar{} for the foreseeable future, targeted observations will be required to recover the transit ephemeris. Successful detection of a second transit of \thisstarb{} would enable further study of this comparatively long-period Warm Jupiter.

\begin{acknowledgments}

We thank the anonymous referee for their comments that have helped to improve this work.
TRB acknowledges support from the Australian Research Council (FL220100117).
This paper includes data collected by the TESS mission that are publicly available from the Mikulski Archive for Space Telescopes (MAST). The specific observations analyzed can be accessed via \dataset[doi:10.17909/qw2d-d851]{https://doi.org/10.17909/qw2d-d851}. Funding for the TESS mission is provided by NASA's Science Mission Directorate. We acknowledge the use of public TESS data from pipelines at the TESS Science Office and at the TESS Science Processing Operations Center.
This research has made use of the Exoplanet Follow-up Observation Program website and the NASA Exoplanet Archive, which are operated by the California Institute of Technology, under contract with the National Aeronautics and Space Administration under the Exoplanet Exploration Program.
This research has made use of the SIMBAD database and VizieR catalogue access tool, operated at CDS, Strasbourg, France. This research has made use of NASA's Astrophysics Data System.
This work has made use of data from the European Space Agency (ESA) mission {\it Gaia} (\url{https://www.cosmos.esa.int/gaia}), processed by the {\it Gaia} Data Processing and Analysis Consortium (DPAC, \url{https://www.cosmos.esa.int/web/gaia/dpac/consortium}). Funding for the DPAC has been provided by national institutions, in particular the institutions participating in the {\it Gaia} Multilateral Agreement.

\end{acknowledgments}

\begin{contribution}



AVe coordinated RV observations of the target, led the analysis, and prepared the manuscript. RAW, CXH, and AVa acted as advisors with regards to the development of the scientific analysis by AVe. MH identified the TESS transit. TRB contributed the asteroseismic analysis. SG advised on the interpretation of the results. AVa, KAC, SJ, RPB, JDC, JYG, DO, MSTP, JT, SS, and SV contributed to the PFS observations. DWL, AB, and SNQ contributed to the TRES observations. RAW, TRF, TRB, JH, SRK, DW, and GZ contributed to the MINERVA-Australis observations. SBH contributed the Zorro observation.

\end{contribution}

%

\vspace{5mm}
\facilities{TESS, \textit{Gaia}, Magellan:Clay (PFS), FLWO:1.5m (TRES), MINERVA-Australis, Gemini:South (Zorro)}


\software{\texttt{emcee} \citep{emcee}
          \texttt{batman} \citep{batman},
          \texttt{isochrones} \citep{isochrones},
           \texttt{kiauhoku} \citep{Tayar2022},
          \texttt{astropy} \citep{astropyI, astropyII, astropyIII}
          }


\clearpage
\appendix

\section{Radial Velocity Data} \label{appendix:RV}

In Table~\ref{tab:RV} we present the RV data used in this work.

\startlongtable
\begin{deluxetable}{lrrr}
\label{tab:RV}
\centering
\tablecaption{Radial velocity data of \thisstar{} used in this work.}
\tablehead{\colhead{BJD} & \colhead{RV (\ms{})} & \colhead{RV error (\ms{})} & \colhead{Instrument}}
\startdata
2460459.92401 & -12.26 & 1.86 & PFS \\
2460464.92556 & -16.12 & 1.79 & PFS \\
2460488.92275 & 20.54 & 2.06 & PFS \\
2460509.86393 & 20.12 & 1.76 & PFS \\
2460515.84542 & -1.59 & 2.07 & PFS \\
2460536.78231 & 14.42 & 2.64 & PFS \\
2460543.81030 & 37.21 & 3.04 & PFS \\
2460603.64357 & 26.28 & 2.65 & PFS \\
2460605.59808 & 22.55 & 2.64 & PFS \\
2460656.55428 & 24.89 & 2.84 & PFS \\
2460836.91424 & 26.39 & 1.83 & PFS \\
2460859.90266 & -22.28 & 1.83 & PFS \\
2460861.87462 & -20.72 & 2.09 & PFS \\
2460873.89481 & 0.33 & 1.82 & PFS \\
2460923.79987 & -16.54 & 2.71 & PFS \\
2460976.64637 & -31.09 & 2.62 & PFS \\
2460452.28436 & 10983 & 28 & MINERVA-Australis \\
2460473.22040 & 10961 & 31 & MINERVA-Australis \\
2460514.20705 & 10975 & 23 & MINERVA-Australis \\
2460525.08566 & 11005 & 23 & MINERVA-Australis \\
2460531.08436 & 10969 & 32 & MINERVA-Australis \\
2460532.12309 & 11061 & 24 & MINERVA-Australis \\
2460533.27580 & 10983 & 68 & MINERVA-Australis \\
2460539.04727 & 11022 & 29 & MINERVA-Australis \\
2460540.04538 & 11012 & 41 & MINERVA-Australis \\
2460544.08762 & 11053 & 46 & MINERVA-Australis \\
2460545.03954 & 11027 & 30 & MINERVA-Australis \\
2460546.08712 & 11037 & 24 & MINERVA-Australis \\
2460547.02725 & 10993 & 45 & MINERVA-Australis \\
2460548.08633 & 11021 & 31 & MINERVA-Australis \\
2460549.02195 & 10980 & 36 & MINERVA-Australis \\
2460553.23839 & 10991 & 97 & MINERVA-Australis \\
2460563.10178 & 11017 & 38 & MINERVA-Australis \\
2460547.9761 & -23 & 17 & TRES \\
2460556.9244 & -21 & 15 & TRES \\
2460560.8755 & -23 & 18 & TRES \\
2460563.8771 & -11 & 13 & TRES \\
2460566.9037 & -48 & 15 & TRES \\
2460575.9538 & -65 & 12 & TRES \\
2460578.9546 & -68 & 13 & TRES \\
2460587.8578 & -46 & 15 & TRES \\
2460591.8491 & -62 & 14 & TRES \\
2460595.8707 & -69 & 15 & TRES \\
2460602.9106 & 3 & 27 & TRES \\
2460607.7573 & -43 & 12 & TRES \\
2460613.8792 & -37 & 15 & TRES \\
2460619.8749 & -21 & 14 & TRES \\
2460624.8468 & -42 & 13 & TRES \\
2460632.8149 & -54 & 16 & TRES \\
2460640.5670 & -89 & 15 & TRES \\
2460646.6811 & -71 & 13 & TRES \\
2460652.7061 & -38 & 12 & TRES \\
2460657.5754 & -35 & 14 & TRES \\
2460662.5882 & -42 & 12 & TRES \\
2460668.6585 & -40 & 16 & TRES \\
2460670.7294 & -43 & 14 & TRES \\
2460678.6501 & -59 & 14 & TRES \\
2460683.5648 & -61 & 15 & TRES \\
2460689.5987 & -82 & 15 & TRES \\
2460694.6192 & -49 & 15 & TRES \\
2460709.5804 & -37 & 11 & TRES \\
2460718.5772 & 0 & 15 & TRES \\
2461050.5873 & -23 & 17 & TRES \\
2461066.5911 & -45 & 12 & TRES \\
2461073.5905 & -47 & 13 & TRES \\
\enddata
\end{deluxetable}

\clearpage
\section{Speckle Imaging} \label{appendix:speckle}

\begin{figure}[h!]
    \centering
    \includegraphics[width=0.5\linewidth]{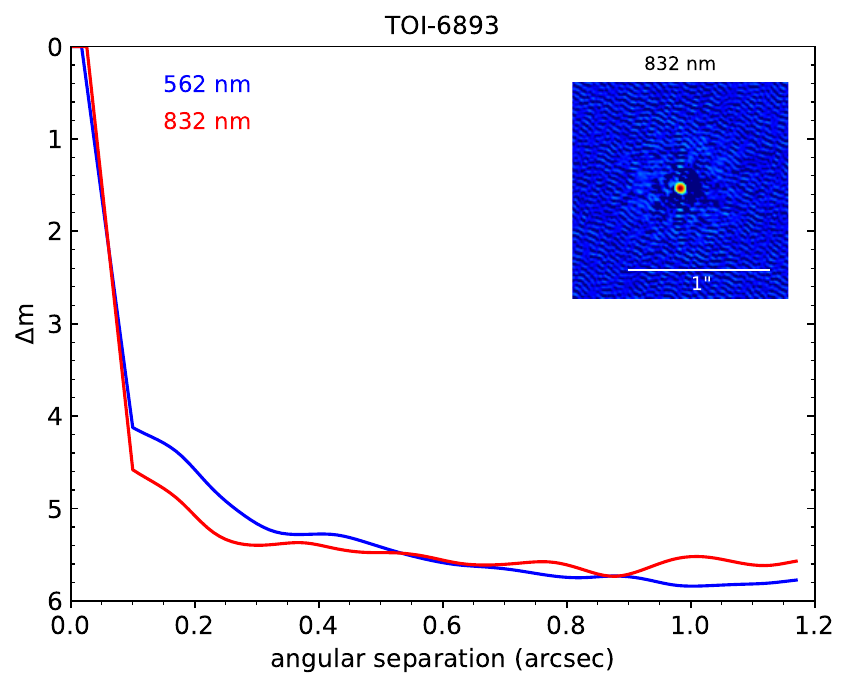}
    \caption{Zorro speckle imaging observations of \thisstar{} obtained on 2024-07-14. The red and blue contrast curves represent the 5$\sigma$ contrast limits as a function of projected separation at 832~nm and 562~nm respectively. The inset figure shows the reconstructed 832~nm image. No additional sources are detected.}
    \label{fig:speckle}
\end{figure}

In Figure~\ref{fig:speckle} we present the Zorro speckle imaging observations of \thisstar{} discussed in Section~\ref{subsec:imaging}.  Observations were taken in the 562~nm and 832~nm filters on 2024-07-14. No additional sources were detected above the 5$\sigma$ contrast limits.

\bibliography{bib}{}
\bibliographystyle{mnras}



\end{document}